\documentclass[sigconf]{acmart}
\AtBeginDocument{%
  }

\setcopyright{none}
\copyrightyear{2026}
\acmYear{2026}
\acmConference[ICCAD '26]{IEEE/ACM International Conference on Computer-Aided Design}{November 08--12, 2026}{San Jose, CA, USA}
\acmBooktitle{}
\acmDOI{}
\acmISBN{}
\renewcommand\footnotetextcopyrightpermission[1]{}

\newcommand{\arxivnotice}{%
  \copyright~2026 Copyright held by the owner/author(s).
  This is the author's version of the work. It is posted here for your personal use.
  Not for redistribution. The definitive Version of Record was published in
  \emph{IEEE/ACM International Conference on Computer-Aided Design (ICCAD '26)},
  November 08--12, 2026, San Jose, CA, USA,
  \url{https://doi.org/10.1145/3831252.3834112}.}

\usepackage{amsmath}
\usepackage{listings}
\usepackage{algorithm,algpseudocode}
\usepackage{graphicx}     % 이미지를 사용하기 위해
\usepackage{subcaption}   % subfigure 환경을 사용하기 위해
\usepackage{makecell}
\usepackage{multirow}
\begin{document}

\title{FlowTT: Exploiting Computation Flow Reuse in Irregular Tensor-Train Embedding}

\author{Jongmin Seok}
\orcid{0009-0003-5037-7252}
\affiliation{
  \institution{Department of AI Semiconductor Engineering\\ Hanyang University}
  \city{Seoul}
  \country{Republic of Korea}
}
\email{seokjongmin@hanyang.ac.kr}

\author{Chae Eun Rhee}
\orcid{0000-0002-7851-1703}
\affiliation{
  \institution{Department of Electronic Engineering\\ Hanyang University}
  \city{Seoul}
  \country{Republic of Korea}
}
\email{crhee@hanyang.ac.kr}
%% \renewcommand{\shortauthors}{Trovato et al.}

\begin{abstract}
Tensor-Train (TT) decomposition effectively compresses large embedding tables in recommendation models, but TT-based embedding lookup remains inefficient because partially shared computation flows across input indices are not fully reused and intermediate results are repeatedly materialized off-chip between sequential TT-core contractions. We present FlowTT, a flow-aware GPU execution framework that reformulates TT gather as a set of prefix-shared irregular computation flows. FlowTT combines flow-aligned prefix-based index grouping, a fused TT-embedding execution path with on-chip intermediate retention, and persistent-thread scheduling with chunk-based work stealing and L2 checkpointing to preserve reuse under skewed workloads. By co-designing task formation, data buffering, and scheduling with the structure of TT gather, FlowTT reduces redundant TT-core operations, global-memory traffic, and load imbalance. On Meta's synthetic recommendation benchmarks (Meta-240, Meta-480, and Meta-788), FlowTT consistently achieves the lowest latency compared to existing methods. At batch size $32{,}768$, it reduces latency by up to $42.2\%$ in inference and $49.2\%$ in training relative to EcoRec, while also achieving the lowest inference peak memory usage. These results show that exposing prefix-shared computation is key to efficient TT-based embedding execution.
% Tensor-Train Decomposition (TTD) effectively compresses large embedding
% tables into compact TT-cores, but TT-based embedding lookup remains
% inefficient because partially shared computation paths across input indices
% are not explicitly reused, and intermediate results are repeatedly
% materialized off-chip between sequential contractions. We present
% \textbf{FlowTT}, a flow-aware GPU execution framework that reformulates
% TT-gather as a set of prefix-shared irregular computation flows.
% FlowTT combines prefix-based unique index selection, a fused
% TT-embedding kernel that retains reusable intermediates in shared memory,
% and persistent-thread scheduling with chunk-based work-stealing and
% L2-mediated checkpoint sharing to balance irregular workloads while
% preserving reuse opportunities. By aligning task formation, on-chip
% buffering, and dynamic scheduling with the structure of TT-gather,
% FlowTT reduces redundant TT-core operations and minimizes off-chip
% data movement. Preliminary evaluation on Meta's synthetic recommendation
% datasets shows that FlowTT consistently achieves lower embedding lookup
% latency and lower peak memory usage than FBTT, EL-Rec, and EcoRec under
% matched TT-rank and batch-size settings. These results suggest that
% computation-flow-aware execution is a promising direction for efficient
% TT-based embedding in modern recommendation systems.
\end{abstract}



%\ccsdesc[500]{Do Not Use This Code~Generate the Correct Terms for Your Paper}
%\ccsdesc[300]{Do Not Use This Code~Generate the Correct Terms for Your Paper}
%\ccsdesc{Do Not Use This Code~Generate the Correct Terms for Your Paper}
%\ccsdesc[100]{Do Not Use This Code~Generate the Correct Terms for Your Paper}


\maketitle
\begingroup
\renewcommand\thefootnote{}%
\footnotetext{\arxivnotice}%
\addtocounter{footnote}{-1}%
\endgroup

\section{Introduction}
\noindent
Recommendation systems are a core technology for delivering personalized content in large-scale online services. In recent years, models based on deep neural networks (DNNs) have become the dominant approach beyond traditional collaborative filtering. Among them, the Deep Learning Recommendation Model (DLRM)~\cite{naumov2019deep} is widely used as a representative architecture that processes sparse features and dense features through embedding tables and Multi-Layer Perceptrons (MLPs), respectively, and then combines them to predict click-through rate (CTR). However, DLRM-based models inherently exhibit clear system bottlenecks. While MLPs are compute intensive, embedding tables are extremely memory intensive, and in production environments their size can reach several terabytes, far exceeding the memory capacity of a single graphics processing unit (GPU). As a result, embedding lookup dominates both training and inference, causing high memory bandwidth demand and frequent off-chip data movement. Distributed execution across multiple GPUs can mitigate this issue, but additional communication overhead from embedding table partitioning and imbalanced memory access still limits overall system scalability~\cite{mudigere2022software, acun2021understanding, agarwal2023bagpipe, zha2022autoshard}.

To reduce the memory footprint of embedding tables, prior work has explored embedding compression techniques~\cite{ginart2021mixed, desai2022trade} such as compositional embeddings~\cite{shi2020compositional} and Tensor-Train decomposition (TTD)~\cite{oseledets2011tensor, yin2021tt}. In particular, TTD represents an embedding table with several low-dimensional TT cores, providing both high compression and strong expressive power~\cite{yin2021tt}. In TTD-based embedding lookup, the retrieved results from multiple TT cores must be combined through sequential matrix operations, a process commonly referred to as TT gather. Unlike a simple lookup, TT gather repeatedly performs the same intermediate computations even when some computation paths are shared across inputs. Because of dependencies across stages, intermediate results are also moved back and forth between stages through off-chip memory. Existing acceleration methods~\cite{wang2022rec, wang2024accelerating} mainly focus on consecutive general matrix multiplication (GEMM) execution based on the CUDA Basic Linear Algebra Subprograms library (cuBLAS)~\cite{nvidia_cublas}, or on keeping intermediate results in on-chip memory as in FlashAttention~\cite{dao2022flashattention, dao2024flashattention2, shah2024flashattention} and FlashGEMM~\cite{zhang2025flashgemm} to maximize data reuse. However, these approaches are mainly designed for regular and uniform execution flows, so they do not fully capture the irregular nature of TT gather, where the computation path changes across inputs. As a result, they miss potential data reuse opportunities and incur unnecessary memory traffic and execution inefficiency.

In this paper, we propose FlowTT to address the structural bottlenecks of TT-based embedding lookup. FlowTT reinterprets TT gather from the perspective of a computation flow shared across inputs and uses this view to reorganize both computation and data movement. The key idea is to group common operations along partially overlapping computation paths and keep them in a reusable form in on-chip memory. This design transforms TT gather from a sequence of independent GEMMs into an on-chip pipeline that follows a shared computation flow. FlowTT also considers the irregular execution patterns caused by the input distribution and dynamically schedules computation flow units to achieve both computation reuse and load balance. This flow-aware execution reduces unnecessary off-chip data movement and maps the input-dependent computation structure more efficiently onto the GPU memory hierarchy. As a result, FlowTT significantly reduces embedding lookup latency and lowers peak memory usage in inference compared with prior TT-based embedding methods~\cite{yin2021tt, wang2022rec, wang2024accelerating}. The main contributions of this paper are as follows.
\begin{itemize}
\item \textbf{Flow-aware execution model.} We redefine TT gather as a computation flow that is partially shared across inputs and introduce a new execution model that uses this flow as the scheduling unit.
\item \textbf{Flow-aligned on-chip execution.} We reorganize computation along the shared computation flow and execute it as a single on-chip pipeline, which minimizes redundant computation and off-chip data movement.
\item \textbf{Dynamic load balancing for irregular flows.} We maintain high resource utilization under input-dependent execution flows through dynamic scheduling at the flow level, which alleviates workload imbalance.
\end{itemize}

\section{Related Work}
\noindent
\subsection{Baseline Tensor-Train Embedding Model}
\begin{figure}[!ht]
    \centerline{\includegraphics[width=0.8\columnwidth]{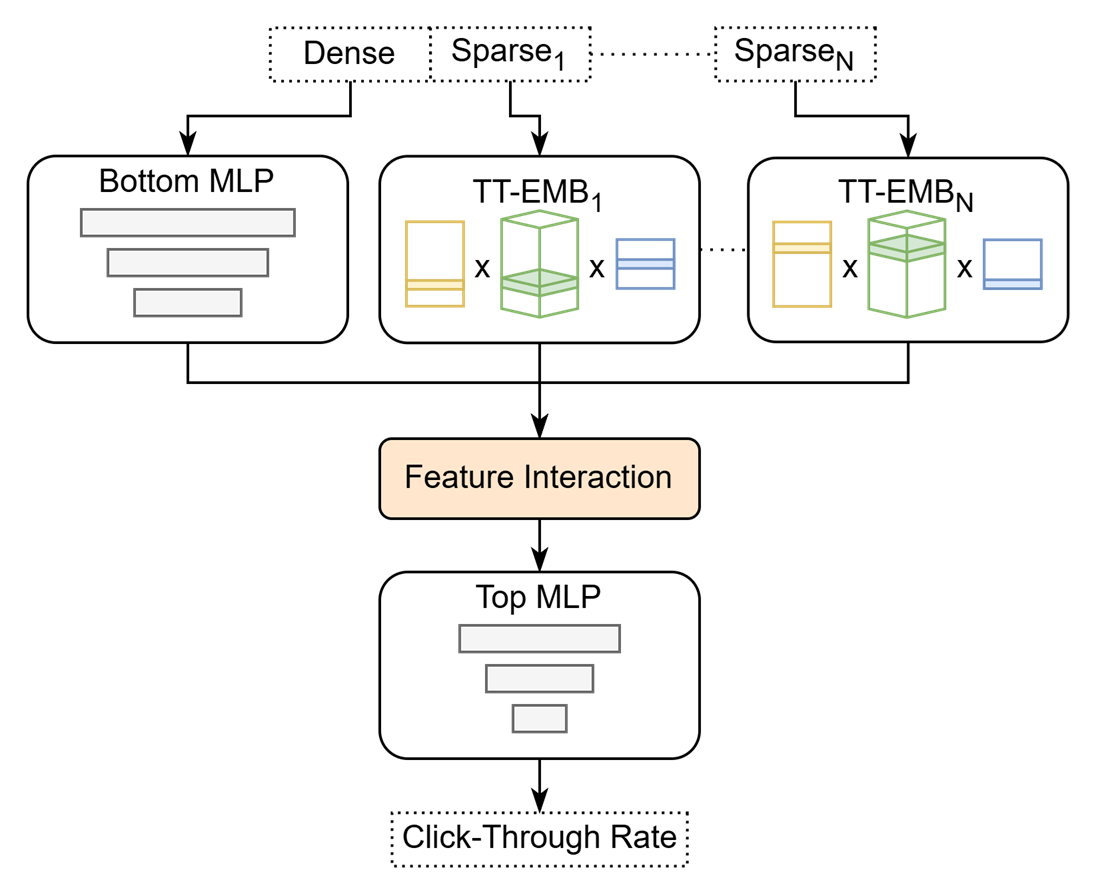}}
    \captionsetup{font=small}
    \caption{Overview of TT-Rec, which decomposes embedding tables into TT-cores and reconstructs sparse embeddings through TT-EMB modules before combining them with dense features for CTR prediction.}
    \label{figure_1}
\end{figure}

Figure~\ref{figure_1} shows the overall architecture of the DLRM with TT-Rec~\cite{yin2021tt}. DLRM consists of a bottom MLP that processes dense features and a set of embedding modules for sparse feature fields, denoted as TT-EMB$_1$--TT-EMB$_N$. The sparse embeddings generated by these modules are passed through feature interaction and then sent to a top MLP to predict CTR.
TT-Rec replaces large embedding tables with a TTD structure by storing each TT-EMB as a set of TT cores. As shown in Figure~\ref{figure_1}, one embedding table is represented by three TT cores, TT-core$_1$, TT-core$_2$, and TT-core$_3$, and an input sparse feature is converted through mixed-radix conversion into the corresponding indices for these TT cores. The model then retrieves the required tensors from the TT cores and combines them through sequential GEMM operations to reconstruct the final embedding vector. This process is commonly called TT gather. However, because each intermediate result must be read again before the next operation, the cost of storing intermediate results and moving data off-chip grows as more small GEMMs are chained together. In addition, the irregular access pattern created by Compressed Sparse Row (CSR) indices reduces GPU memory parallelism and causes both kernel launch overhead and memory bandwidth pressure. Therefore, the main bottleneck of TT embedding lies less in the compression itself than in the execution path used to reconstruct the compressed representation.

\subsection{Optimization of Tensor-Train Embedding}
EL-Rec~\cite{wang2022rec} is a framework proposed to apply TT embedding to practical training and inference systems, and it alleviates the structural bottlenecks of TT-Rec at the primitive operator level. It reorders operations across TT cores to enable more contiguous memory access, replaces the standard PyTorch embedding operator with a high-performance compressed embedding operator, and reduces communication and data preparation overhead through feature reordering and pipeline-based asynchronous training. These optimizations substantially improve the practicality of TT embedding. However, EL-Rec primarily organizes reuse through operator-level execution and training-pipeline optimizations. It does not elevate a prefix-shared intermediate into a block-local execution state that is explicitly buffered and reused in shared memory across multiple inputs.

EcoRec~\cite{wang2024accelerating} further reduces redundant computation through operation-pattern reorganization, C-M/M-C contraction patterns, and sorted-index micro-batching. It also improves distributed training scalability by overlapping table-level computation and communication through table-wise pipeline scheduling. This design is highly effective for reducing TT-pair redundancy and communication bottlenecks in distributed training. The distinction of FlowTT is not that EcoRec ignores redundancy, but that EcoRec exposes reuse mainly at the TT-pair, micro-batch, and table-pipeline levels. In contrast, FlowTT brings prefix-shared reuse into the GPU kernel itself by treating the shared intermediate as a GPU block/shared-memory execution state.

Taken together, EL-Rec and EcoRec significantly improve the applicability and scalability of TT embedding in single-GPU and distributed settings, respectively. FlowTT is complementary to these efforts, but targets a finer execution granularity. Rather than only improving operator ordering, access locality, micro-batching, or pipeline scheduling, FlowTT defines TT gather as a prefix-shared computation flow and maps each reusable prefix state directly onto GPU block-level execution and shared-memory buffering.

\subsection{On-Chip Pipeline and Scheduling Techniques}
System- and architecture-level techniques for improving the efficiency of consecutive matrix operations have been studied extensively. FlashGEMM~\cite{zhang2025flashgemm} is a representative approach on x86 processors that combines loop fusion, cache-resident data reuse, and Vector Neural Network Instructions (VNNI) to maximize reuse across consecutive GEMMs. FlashAttention~\cite{dao2022flashattention, dao2024flashattention2, shah2024flashattention} reorganizes multiple stages of the attention computation into a single on-chip pipeline and keeps intermediate results in shared memory and registers, which effectively reduces dynamic random access memory (DRAM) spill. These studies~\cite{dao2022flashattention, dao2024flashattention2, shah2024flashattention, zhang2025flashgemm} share a common design principle. They integrate consecutive operations into one pipeline and minimize the need to store intermediate results between stages, thereby maximizing data reuse in fast memory levels through loop fusion, memory locality optimization, and on-chip buffers. However, these techniques are generally designed for relatively static and uniform dense pipelines, where it is known in advance which operations will be combined and where intermediate results will be reused. That assumption does not hold directly for computation flows whose paths change across inputs.

%From the scheduling perspective, the persistent thread (PT) model and work stealing are widely used approaches for irregular workloads. 

From the scheduling perspective, persistent threads (PT) and dynamic task stealing have long been studied as practical mechanisms for irregular GPU workloads~\cite{gupta2012study, tzeng2010task, aila2009understanding, chatterjee2011dynamic}. PT keeps a small number of resident thread blocks on each streaming multiprocessor (SM) and fetches work dynamically~\cite{gupta2012study, aila2009understanding}, while task stealing allows idle workers to take unfinished work and mitigate load imbalance~\cite{tzeng2010task, chatterjee2011dynamic}. These techniques are broadly useful as general scheduling strategies for irregular workloads, but they do not by themselves identify which intermediate results should be reused or how such reuse should be mapped to shared-memory-resident execution states.

\subsection{Problem Definition}
The core operation in TT-based embedding lookup is TT gather, which takes an index converted by mixed-radix conversion into $(d^{(1)}, d^{(2)}, d^{(3)})$, retrieves the corresponding slice from each TT core, and combines them through sequential contractions to reconstruct the embedding vector. If two different inputs share the same $(d^{(1)}, d^{(2)})$ and differ only in the last digit $d^{(3)}$, then the intermediate result produced after contracting the first two TT cores is identical, while only the final stage differs. TT gather is therefore neither a set of fully independent dense GEMM chains nor a simple repetition of identical duplicate lookups. Instead, the full input set forms an irregular computation flow that combines shared prefix computation with input-specific suffix computation.

This point distinguishes our problem definition from prior TT embedding acceleration studies. EL-Rec and EcoRec already exploit reuse or redundancy in TT embedding through operator reordering, TT-pair contraction, sorted micro-batching, table-wise scheduling, and distributed overlap. However, their reusable units are mainly organized at the operator, TT-pair, micro-batch, or table level. FlowTT instead promotes the prefix-shared intermediate itself to a GPU execution unit, so that multiple inputs sharing the same prefix can reuse the same intermediate directly inside a block through shared-memory buffering. By contrast, FlashAttention and FlashGEMM provide strong examples of on-chip reuse and fusion of consecutive operations, but they assume dense pipelines with static data flow and relatively uniform tile structure. In an irregular operator such as TT gather, where the execution path changes with the input and only some stages are shared, simply fusing consecutive GEMMs does not determine which intermediate should be reused or when it should be reused, and it also does not resolve the load imbalance caused by skewed prefix frequencies.

Accordingly, the problem we address is more specific than simple GEMM fusion or improved memory locality. We must identify the partial computations that can be shared across inputs based on mixed-radix prefixes and reconstruct them as GPU block-level execution units. We must keep these prefix-shared intermediates on a shared-memory-centered execution path to avoid repeated off-chip materialization of per-input intermediates. We must also combine this design with dynamic scheduling so that high SM utilization is maintained even under irregular workloads with highly skewed prefix frequencies. In this sense, the core of our fused TT embedding kernel is not merely to merge consecutive GEMMs into one kernel. It is to redefine TT gather as a prefix-aware irregular computation problem and to co-design task formation, intermediate buffering, and load balancing accordingly.

\section{Proposed Method}
\noindent
\subsection{Overview}
\begin{figure}[htbp]
    \centering
    \begin{subfigure}[b]{0.45\linewidth}
        \centering
        \includegraphics[width=\linewidth]{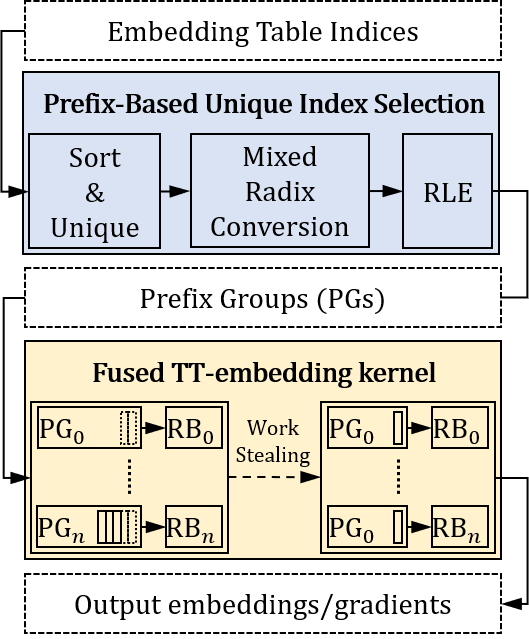}
        \caption{}
        \label{fig:FlowTT_overview_a}
    \end{subfigure}
    \hfill
    \begin{subfigure}[b]{0.45\linewidth}
        \centering
        \includegraphics[width=\linewidth]{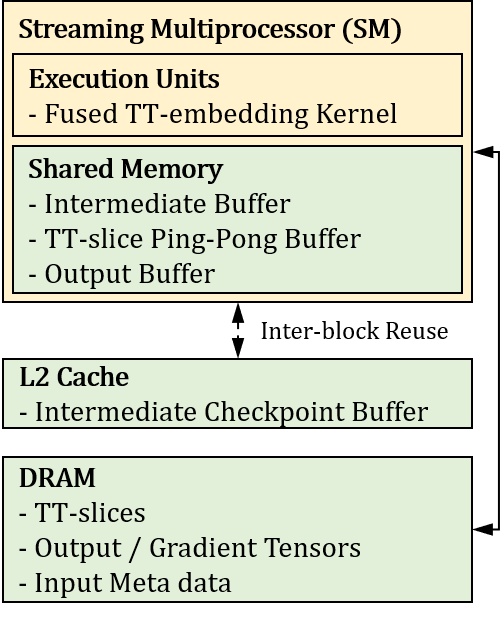}
        \caption{}
        \label{fig:FlowTT_overview_b}
    \end{subfigure}
    \caption{Overview of FlowTT. (a) Prefix-based index selection groups input indices into prefix-level tasks via sorting, mixed-radix conversion, and RLE, enabling reuse and dynamic load balancing through work stealing. (b) Execution and memory hierarchy where each SM runs the fused TT-embedding kernel with shared-memory-based intermediate reuse, while L2 cache enables inter-block reuse and DRAM stores TT slices and tensors.}
    \label{fig:FlowTT_overview}
    \label{fig:overview}
\end{figure}

Figure~\ref{fig:FlowTT_overview} presents the overall structure of FlowTT through two views: prefix-based task formation and the execution and memory hierarchy.
As shown in Figure~\ref{fig:FlowTT_overview_a}, input embedding indices are grouped by shared high-order TT core indices through Sort and Unique, Mixed-radix Conversion, and Run-Length Encoding (RLE). Each prefix group (PG) represents inputs that share a partial computation path, forming a natural execution unit that removes redundant TT core operations. These groups are assigned to resident blocks in a persistent-thread style. Each block first processes its own group and, if it finishes early, steals remaining chunks from unfinished groups, preserving prefix-based reuse while reducing load imbalance.
Figure~\ref{fig:FlowTT_overview_b} shows the execution and memory hierarchy on the GPU. Each resident block (RB) on an SM runs a fused TT-embedding kernel and maximizes intermediate reuse through shared memory, which contains an intermediate buffer, a TT-slice ping-pong buffer for overlapping prefetch and computation, and an output buffer. TT slices, input indices, and output or gradient tensors reside in DRAM, while only required data is loaded on-chip. For inputs sharing the same prefix, intermediate results remain in shared memory and are reused, eliminating off-chip storage of intermediates.
When a PG is processed by multiple blocks due to work stealing, inter-block reuse may be broken. To address this, FlowTT selectively stores reusable intermediates in checkpoint buffers in the level 2 (L2) cache, allowing other blocks to reload and reuse them. This design combines block-local reuse in shared memory with inter-block reuse via L2 cache, maintaining high efficiency under irregular workloads.
Overall, FlowTT makes computation flow explicit through prefix-based task formation, maximizes reuse via shared-memory-centric on-chip execution, and combines persistent threads, work stealing, and L2-based sharing to achieve high GPU utilization.

Section~3.2 describes flow-aware fused TT-embedding execution. Section~3.3 presents prefix-based index grouping. Section~3.4 explains dynamic load balancing. Section~3.5 describes inter-block intermediate sharing through the L2 cache.

\subsection{Flow-Aware Fused TT Embedding Execution}
FlowTT uses fused TT-embedding kernels for the forward and backward paths, respectively. Each fused kernel combines sequential TT-core contractions and aggregation steps to reduce intermediate materialization.
It reorganizes the computation so that operations sharing the same intermediate result are grouped into one execution unit, allowing each intermediate to be computed once and reused. This design maximizes intermediate reuse and reduces both unnecessary memory access and redundant computation.

\begin{figure}[htbp]
  \centering
  \includegraphics[width=\linewidth]{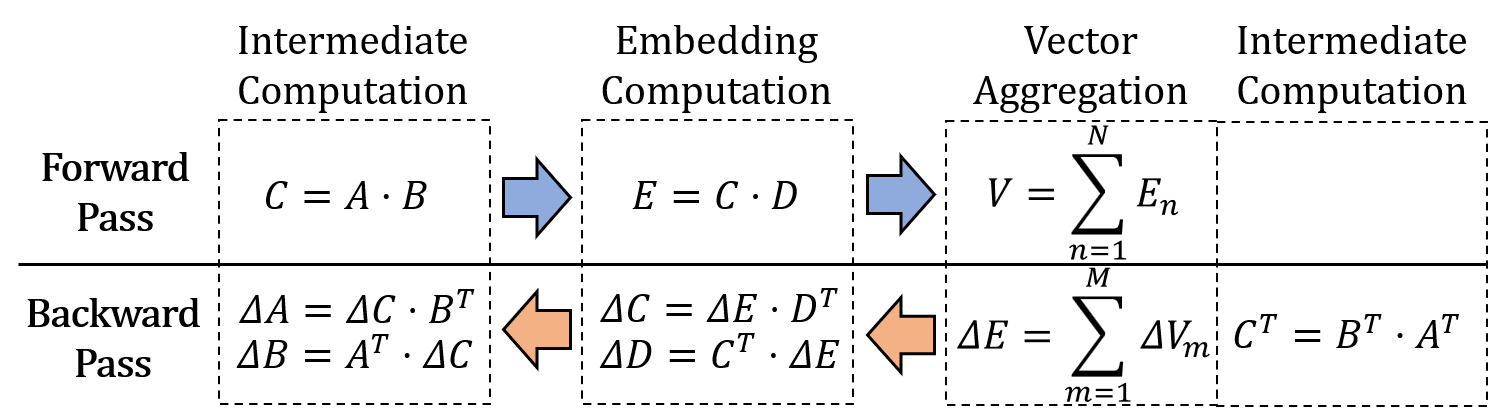}
  \caption{Forward and backward computation in TT embedding. The matrices $A$, $B$, and $D$ denote the TT slices retrieved from TT-core$_1$, TT-core$_2$, and TT-core$_3$, respectively.}
  \label{fig:forward_backward}
\end{figure}

As illustrated in Figure~\ref{fig:forward_backward}, the overall computation consists of three stages: \textit{Intermediate Computation}, \textit{Embedding Computation}, and \textit{Vector Aggregation}. Both the forward pass and the backward pass follow this structure. In the forward pass, the kernel first computes the intermediate $C=A\cdot B$ from the TT slices $A$ and $B$ taken from TT-core$_1$ and TT-core$_2$. It then reconstructs the embedding vector $E=C\cdot D$ by combining $C$ with the TT slice $D$ from TT-core$_3$, and finally produces the output $V$ through pooling.
The backward pass proceeds in the reverse order. After obtaining $\Delta E$ in the aggregation stage, the kernel computes $\Delta D=C^T\cdot \Delta E$ and $\Delta C=\Delta E\cdot D^T$, and then derives $\Delta A$ and $\Delta B$ from $\Delta C$. When multiple outputs share the same intermediate $C$, the corresponding $\Delta C$ values can be accumulated first and then used to compute the gradients. This property is summarized in Equation~(\ref{eqn:aggregation_gradient}).
\begin{align} \label{eqn:aggregation_gradient}
    \begin{split}
    \Delta A &= \sum_{n=1}^{N}{\Delta A_{n}} = \sum_{n=1}^{N}{\Delta C_{n} \cdot B^T} = \Bigg(\sum_{n=1}^{N}{\Delta C_{n}}\Bigg) \cdot B^T \\
    \Delta B &= \sum_{n=1}^{N}{\Delta B_{n}}  = \sum_{n=1}^{N}{A^T \cdot \Delta C_{n}} = A^T \cdot \Bigg(\sum_{n=1}^{N}{\Delta C_{n}}\Bigg)\\
    \end{split}
\end{align}

Assume that \(N\) outputs share the same intermediate \(C\). The fused TT-embedding kernel incorporates several optimizations to improve both memory and compute efficiency. These optimizations focus on on-chip intermediate retention, overlap of prefetch and computation, and efficient aggregation.

\textbf{On-Chip Intermediate Retention.} By default, FlowTT keeps intermediate results in shared memory and avoids writing them to global memory. Only the reusable intermediate results are stored for inter-block reuse, as illustrated in Figure~\ref{fig:FlowTT_overview_b}. Conventional TT-embedding kernels store intermediate results in global memory and subsequently reload them, incurring additional latency and increasing peak memory usage. FlowTT eliminates this off-chip traffic through kernel fusion, thereby reducing both latency and memory footprint. In addition, the kernel checks whether the intermediate value already stored in shared memory matches the one required for the current computation. When a match is found, the existing intermediate is reused, avoiding redundant computation.

\textbf{Prefetch and Compute Overlap.} FlowTT employs a ping-pong buffer-based prefetch mechanism. Shared memory is partitioned into a region for current computation and another for preloading data for subsequent computation. Asynchronous data transfers between global memory and shared memory enable overlap between TT-slice loading and computation. Since the middle TT core is typically larger due to its two rank dimensions, FlowTT distributes this computation across multiple blocks to reduce shared memory usage while maintaining effective overlap between prefetch and computation.

\textbf{Efficient Aggregation.} FlowTT reduces the reliance on atomic operations during embedding vector pooling and gradient aggregation. Issuing atomic operations for every vector can lead to serialization and increased thread synchronization overhead, thereby degrading SM utilization. To mitigate this issue, FlowTT performs aggregation in batches across multiple embedding vectors and uses a shared-memory output buffer to store intermediate aggregation results. This design lowers the frequency of atomic operations, reduces thread idle time, and improves overall SM utilization.
% \textbf{Fused Weight Update.}
% FlowTT fuses the weight update into the backward phase. During backward execution, gradient contributions for each TT-core slice are aggregated once and stored in global memory. The update then directly uses the aggregated gradient and applies it.
% This removes optimizer side aggregation, reduces global memory traffic, and lowers update overhead.

\subsection{Flow-Aligned Prefix-Based Index Grouping}
In this subsection, we describe a prefix-based index selection method that groups inputs sharing the same upper TT-core computations in TT gather. The key idea is to combine indices with identical high-order digits in the mixed-radix representation into a single execution unit, enabling reuse of the intermediate results associated with the corresponding prefix. 
The method first decomposes an embedding index $g$ into per-core indices of each TT core. Given TT-core dimensions $(p_1,\ldots,p_L)$, the index of the $k$th TT core is computed as in Equation~(\ref{eq:mixed_radix}).
\begin{equation}
    \label{eq:mixed_radix}
    d^{(k)} 
    = 
    \left\lfloor 
    \frac{\,g\,}{\displaystyle\prod_{k < j} p_j}
    \right\rfloor
    \bmod p_k
\end{equation}
Here, an index of an earlier TT core corresponds to a higher-order digit. Therefore, multiple input indices can share the same prefix $(d^{(1)}, d^{(2)}, \ldots)$. 
Recommendation system workloads often follow a long-tail distribution~\cite{sethi2022recshard, adnan2021accelerating}, which increases the likelihood of repeated indices and shared high-order prefixes~\cite{wang2024accelerating}.

\begin{figure}[htbp]
  \centering
  \includegraphics[width=0.9\linewidth]{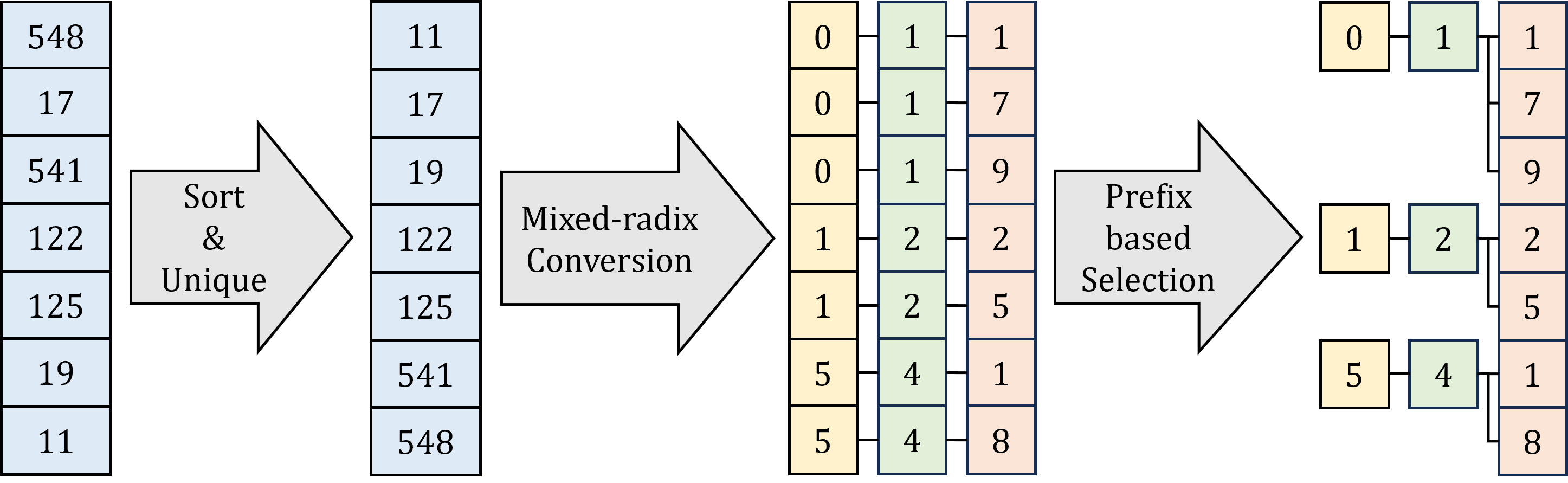}
  \caption{Prefix-Based Unique Index Selection extracts indices that share the same prefix by sorting and deduplicating the indices, applying mixed-radix conversion, and performing run-length encoding.}
  \label{fig:prefix_index}
\end{figure}

Figure~\ref{fig:prefix_index} shows the full grouping process that exploits this shared prefix structure. The input embedding indices are first sorted in ascending order and deduplicated so that redundant computation is removed in advance. The sorted indices are then decomposed into TT core indices $(d^{(1)}, d^{(2)}, \ldots, d^{(L)})$ using Equation~(\ref{eq:mixed_radix}). Because the high-order digits correspond to prefixes, they remain ordered together while preserving the original sorted order. Finally, RLE is applied to the sorted prefixes to extract contiguous ranges of indices that share the same prefix, and each range is treated as one group.
The resulting PGs serve as execution units for the fused TT embedding kernel and are assigned to thread blocks that are later executed as resident blocks under a persistent thread model. Since the indices in the same PG share the TT-core$_1$ and TT-core$_2$ computations, the corresponding intermediate result can be computed only once and then reused. As a result, this method reduces redundant computation and increases the reuse rate of shared-memory-based intermediate results.

\subsection{Dynamic Load Balancing via Persistent Threads and Work Stealing}
\begin{figure}[htbp]
    \centering
    \begin{subfigure}[b]{0.45\linewidth}
        \centering
        \includegraphics[width=0.9\linewidth]{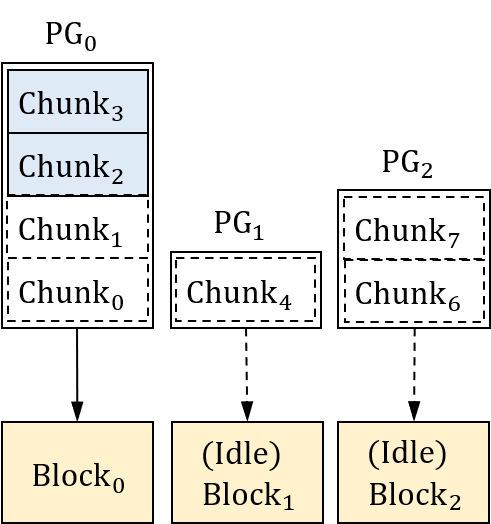}
        \caption{}
        \label{fig:ps_ws_a}
    \end{subfigure}
    \hfill
    \begin{subfigure}[b]{0.45\linewidth}
        \centering
        \includegraphics[width=0.9\linewidth]{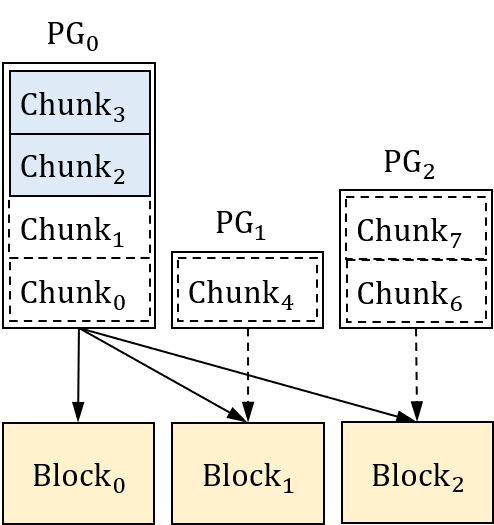}
        \caption{}
        \label{fig:ps_ws_b}
    \end{subfigure}
    \caption{Dynamic load balancing with persistent threads and work stealing. (a) Static prefix mapping leads to load imbalance, where some blocks become idle while others processing long prefix groups become bottlenecks. (b) FlowTT allows each block to process its assigned PG first and then steal remaining chunks from unfinished groups. Chunk-level stealing reduces atomic contention while preserving prefix locality.}
    \label{fig:pt_ws}
\end{figure}

Prefix-based grouping maximizes intermediate reuse across inputs that share the same prefix. However, the length of each PG can vary widely because recommendation system workloads often follow a long-tail distribution~\cite{sethi2022recshard, adnan2021accelerating}. When one block is statically assigned to one PG, the high reuse rate comes at the cost of load imbalance. A few long PGs become stragglers, while other blocks reach an idle state early. As illustrated in Figure~\ref{fig:ps_ws_a}, blocks assigned to short PGs finish quickly, while a block processing a long PG delays overall execution. In the figure, dashed chunks denote completed work, whereas solid chunks represent remaining work.

A simple solution is to split the work into smaller units and let multiple blocks process the same prefix at the same time. However, this substantially reduces block-local reuse of intermediate results. TT gather therefore faces a direct trade-off between load balance and data reuse. To address this issue, FlowTT combines persistent threads with chunk-based work stealing. The sorted prefixes are first divided so that each block receives a similar number of prefixes, and each block prioritizes its assigned PG. Each PG is further divided into multiple chunks, and blocks obtain the next chunk dynamically through an atomic counter.
As shown in Figure~\ref{fig:ps_ws_b}, each block processes the chunks in its own PG sequentially and reuses the same prefix intermediate kept in shared memory. Once a block exhausts its local work, it becomes idle and then continues by taking remaining chunks from another PG that has not yet finished. This chunk-based work stealing redistributes work across blocks to reduce idle time and improve SM utilization, while preserving prefix continuity and therefore the opportunity to reuse intermediate results.

\subsection{Inter-Block Intermediate Sharing via L2 Cache}
\begin{figure}[t]
    \centering
    \begin{subfigure}[b]{0.45\linewidth}
        \centering
        \includegraphics[width=0.9\linewidth]{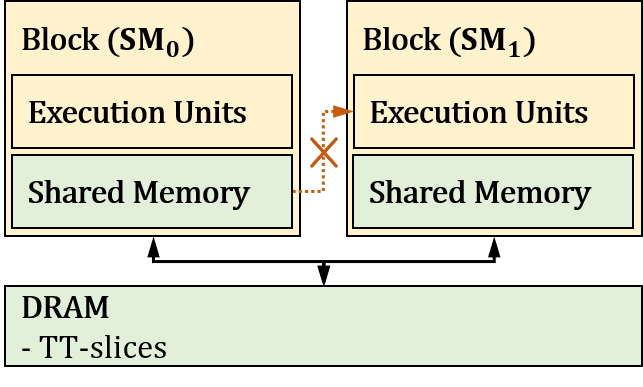}
        \caption{}
        \label{fig:l2_checkpoint_a}
    \end{subfigure}
    \hfill
    \begin{subfigure}[b]{0.45\linewidth}
        \centering
        \includegraphics[width=0.9\linewidth]{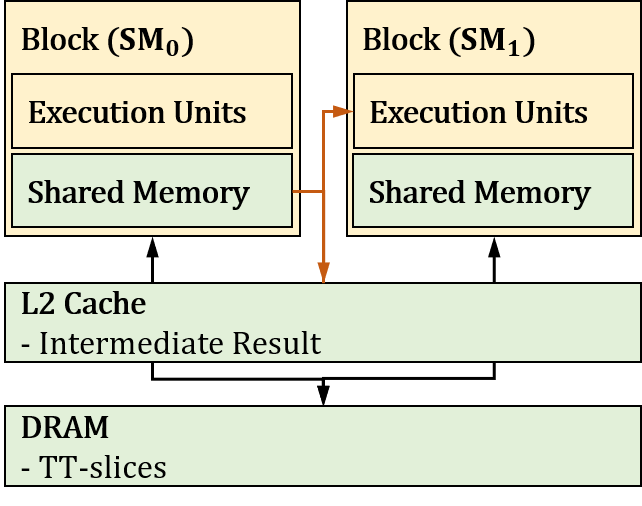}
        \caption{}
        \label{fig:l2_checkpoint_b}
    \end{subfigure}
    \caption{Inter-block intermediate sharing under work stealing. (a) Without L2 checkpointing, a block that steals work cannot access the original block’s shared memory and must recompute prefix-shared intermediates. (b) FlowTT stores reusable intermediates in an L2 checkpoint buffer, allowing a stealing block to reload them into shared memory when the prefix matches and continue suffix computation.}
    \label{fig:l2_checkpoint}
\end{figure}

Chunk-based work stealing is effective for mitigating stragglers, but it breaks reuse of intermediate results across blocks. As shown in Figure~\ref{fig:l2_checkpoint_a}, blocks can share chunks but cannot access one another's shared memory. Therefore, even when a stolen chunk shares the same prefix, the prefix-shared intermediate cannot be reused and must be recomputed. This weakens the benefit of prefix-based reuse.

To address this problem, FlowTT separates block-local reuse from inter-block handoff. Each block continues to reuse intermediate results in shared memory, while only the most recent intermediate with high reuse potential is saved in a checkpoint buffer and kept resident in the L2 cache. Because the L2 cache is shared across blocks, another block can access this intermediate when needed.
As shown in Figure~\ref{fig:l2_checkpoint_b}, a block that steals a chunk loads the intermediate from the L2 cache into shared memory and continues the suffix computation only when the prefix of the stolen chunk matches the checkpoint. Otherwise, it recomputes the intermediate. This selective reuse limits unnecessary L2 accesses while still enabling reuse across blocks. To bound L2 pressure, the number of checkpoints is limited in proportion to the number of resident blocks. 
%FlowTT also avoids storing intermediate results in global memory, which preserves its on-chip reuse oriented design. 
In implementation, we control checkpoint residency in the L2 cache with \texttt{cudaLimitPersistingL2CacheSize}.

\section{Evaluation}
\noindent
% Requires: \usepackage{booktabs}, \usepackage{multirow}
% Paper safe revised evaluation section
% This version reflects the current paper setup, where detailed figure based analyses focus on inference,
% while Table 2 additionally reports representative training results.

%\section{Evaluation}

\subsection{Experimental Setup}

We evaluate FlowTT using the synthetic recommendation dataset released by Meta and compare it against representative TT-based embedding lookup implementations, including FBTT~\cite{facebook_fbtt_embedding}, EL-Rec~\cite{wang2022rec}, and EcoRec~\cite{wang2024accelerating}. For FBTT, the optimized FBTT-Embedding implementation~\cite{facebook_fbtt_embedding} provided by Meta is used as a simple baseline following the standard TT-Rec formulation. 
We intentionally use the same Meta benchmark~\cite{meta_dlrm_datasets_2022} as EcoRec to ensure a direct and fair comparison with the strongest prior baseline. Since our contribution focuses on the execution efficiency of TT-based embedding lookup rather than dataset-specific model design, keeping the benchmark fixed is important for isolating the effect of the proposed method. This matched-benchmark setting controls dataset-dependent factors such as the number of tables, table-size distribution, pooling factors, and index-access reuse patterns, allowing the observed performance differences to be attributed more directly to the execution framework itself. The dataset consists of three subsets from the Meta benchmark: Meta-240, Meta-480, and Meta-788, containing 240, 480, and 788 embedding tables, respectively~\cite{meta_dlrm_datasets_2022, wang2024accelerating}. Each embedding table is decomposed into three TT cores. Unless otherwise specified, the TT ranks are fixed to $r_1 = r_2 = 32$, and the embedding dimension is set to $d = 32$. This configuration follows the standard operating regime of TT embeddings adopted by TT-Rec, EL-Rec, and EcoRec; FlowTT remains the fastest method across embedding dimensions $\{16, 32, 64\}$ and TT ranks $\{8, 16, 32\}$, and the same method ordering is reproduced on a second GPU architecture (NVIDIA RTX 4090), as the per-block L2 checkpoint occupies only about 1\,KB. FlowTT is also numerically equivalent to reference TT lookup implementations, with AUC differences below 0.001.
All results are obtained as averages over multiple iterations after sufficient warm-up.
Experiments are conducted on the server configuration described in Table~\ref{tab:server_setup}. We measure end-to-end latency (including preprocessing), kernel latency, and peak memory usage. To further analyze the contributions of scheduling and prefix-based reuse, we conduct ablation studies.
For batch-size scaling, latency breakdown, and ablation studies, we use the same dataset and TT-rank configuration unless otherwise specified.
%detailed breakdown and reuse analysis, we present inference results, as illustrated in Figure~\ref{fig:latency_per_batch_size}, Figure~\ref{fig:e2e_breakdown} and Figure~\ref{fig:reuse_ablation}, since these figures focus on latency trends. The relative ordering of training latency is consistent with inference, as shown in Table~\ref{tab:lat_mem_compare}. Accordingly, representative training results are included in Table~\ref{tab:lat_mem_compare}, while other analyses are based on inference.

\begin{table}[ht]
    \centering
    \caption{Server configuration used for evaluating Meta-240, Meta-480, and Meta-788 benchmarks.}
    \label{tab:server_setup}
    \small
    \setlength{\tabcolsep}{4pt}
    \begin{tabular}{ll}
        \toprule
        CPU & Intel Xeon Gold 5218R CPU @ 2.10GHz \\
        System memory & 376 GiB DRAM \\
        GPU & NVIDIA RTX A6000 (48 GB, CC 8.6) \\
        Software & 64-bit Linux, CUDA Toolkit 12.8, \\
         & Python 3.11.15, PyTorch 2.10.0\\
        \bottomrule
    \end{tabular}
\end{table}

\subsection{Latency and Memory Comparison under Varying Batch Sizes}

\begin{figure}[ht]
    \centering
    \begin{subfigure}[b]{\columnwidth}
        \centering
        \includegraphics[width=0.9\columnwidth]{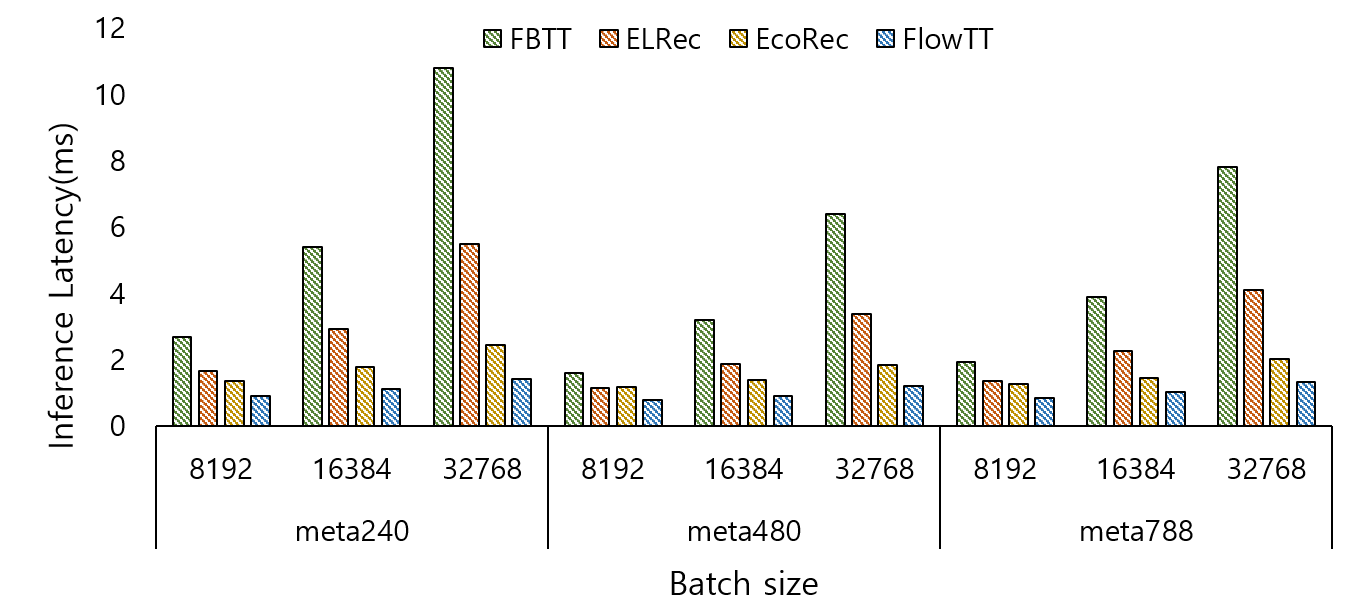}
        \caption{}
        \label{fig:latency_per_batch_size_a}
    \end{subfigure}
    \begin{subfigure}[b]{\columnwidth}
        \centering
        \includegraphics[width=0.9\columnwidth]{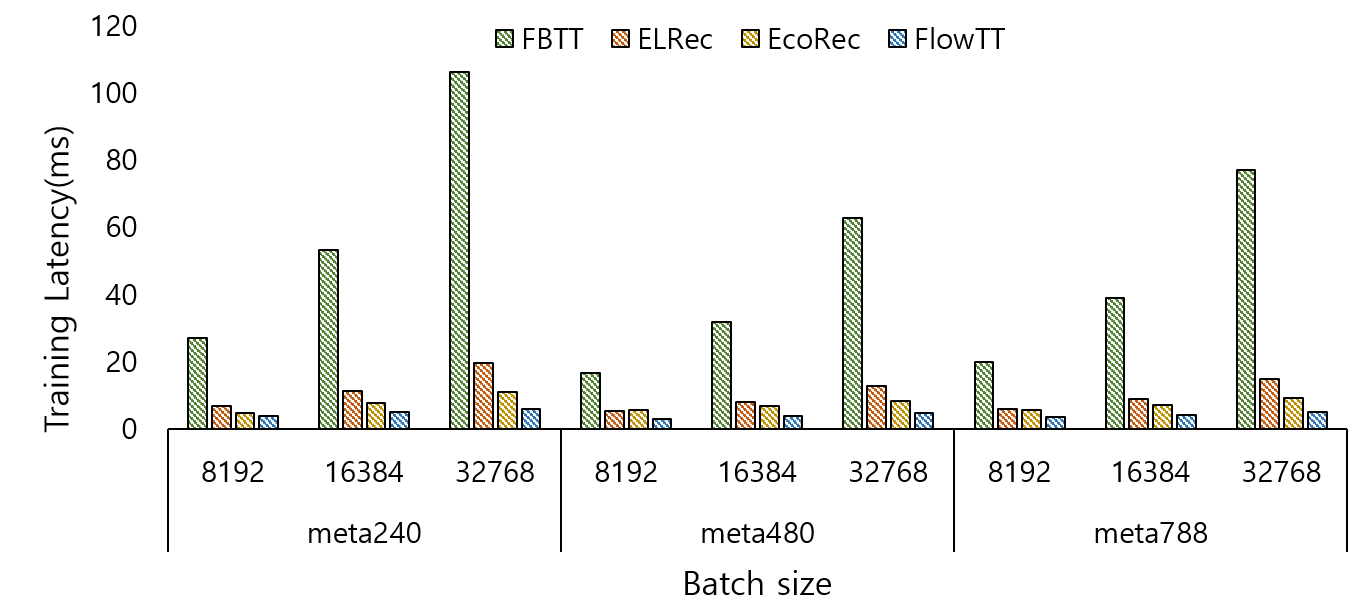}
        \caption{}
        \label{fig:latency_per_batch_size_b}
    \end{subfigure}
    \caption{(a) Inference and (b) Training latency comparison under varying batch sizes across Meta-240, Meta-480, and Meta-788 with fixed TT rank. FlowTT consistently outperforms all baselines, with performance gains becoming more pronounced as the batch size increases.}
    \label{fig:latency_per_batch_size}
\end{figure}

Figure~\ref{fig:latency_per_batch_size} compares latency under varying batch sizes while keeping the dataset and TT rank fixed.
Figure~\ref{fig:latency_per_batch_size_a} shows the inference latency. Across all datasets and batch sizes, FlowTT consistently achieves the lowest latency, and the performance gap with respect to the baselines increases as the batch size grows. This advantage becomes more pronounced in the large-batch regime, where a larger fraction of inputs share common prefixes, increasing opportunities for intermediate reuse. In addition, the fixed overhead of task formation and scheduling is amortized over more inputs, allowing the benefits of FlowTT’s on-chip execution and reuse mechanism to be more fully realized.
Figure~\ref{fig:latency_per_batch_size_b} presents the training latency. A similar trend is observed, where FlowTT consistently outperforms all baselines across different datasets and batch sizes. Notably, the performance gap widens more significantly compared to inference, reflecting the higher computational intensity of training and the greater benefits from eliminating redundant computations.
Overall, this trend is consistently observed across Meta-240, Meta-480, and Meta-788, indicating that the performance benefits of FlowTT generalize across different embedding table scales.

Table~\ref{tab:lat_mem_compare} presents latency and peak memory comparisons for both inference and training under a representative large-batch setting. For inference, FlowTT achieves both the lowest latency and the smallest peak memory usage across all datasets. Compared to the strongest baseline, EcoRec, FlowTT reduces latency by 42.2\%, 33.9\%, and 34.3\% on Meta-240, Meta-480, and Meta-788, respectively. These gains can be attributed to reduced off-chip memory accesses enabled by prefix-based intermediate reuse and on-chip execution.
In training, FlowTT also achieves the lowest latency across all datasets, with latency reductions of 47.4\%, 37.1\%, and 49.2\% compared to EcoRec. This suggests that, similar to inference, intermediate reuse and kernel-level optimizations dominate the overall execution time.
In contrast, while FlowTT attains the lowest peak memory usage in inference, EcoRec shows lower peak memory usage in training. This is due to the additional buffer required by FlowTT for gradient aggregation, indicating a trade-off between memory footprint and latency reduction. Nevertheless, FlowTT consistently provides the best performance in terms of latency.
Beyond embedding-only kernels, FlowTT also achieves the lowest end-to-end DLRM training latency on Meta-788, running at 977\,ms per iteration versus 1{,}166\,ms for EcoRec, and it remains faster than EcoRec on full real recommendation traces (Amazon CDs and Amazon Video Games), where embeddings account for a much smaller fraction of the training step.

\begin{table*}[t]
    \centering
    \caption{Latency and peak memory comparison under a representative large-batch setting across Meta-240, Meta-480, and Meta-788 with fixed TT rank. All methods use three TT cores with $r_1 = r_2 = 32$ and embedding dimension $d = 32$. Lower values are better, and \textbf{bold numbers} indicate the best results.}
    \label{tab:lat_mem_compare}
    \small
    \setlength{\tabcolsep}{4pt}
    \begin{tabular}{l|ccc|cccc|cccc}
        \toprule
        & \multirow{2}{*}{Dataset} & \multirow{2}{*}{Batch} & \multirow{2}{*}{TT rank} &
        \multicolumn{4}{c|}{Latency (ms)$\downarrow$} &
        \multicolumn{4}{c}{Peak Memory (MB)$\downarrow$} \\
        \cmidrule(lr){5-8} \cmidrule(lr){9-12}
        & & & & FBTT\cite{facebook_fbtt_embedding} & EL-Rec\cite{wang2022rec} & EcoRec\cite{wang2024accelerating} & FlowTT & FBTT & EL-Rec & EcoRec & FlowTT \\
        \midrule
        \multirow{3}{*}{Inference}
        & Meta-240 & 32,768 & (32, 32) &10.81 & 5.50 & 2.44 & \textbf{1.41} & 4912.48 & 2657.85 & 2148.95 & \textbf{1970.45} \\
        & Meta-480 & 32,768 & (32, 32) &6.41 & 3.38 & 1.83 & \textbf{1.21} & 6069.71 & 3325.92 & 2697.67 & \textbf{2541.50} \\
        & Meta-788 & 32,768 & (32, 32) &7.82 & 4.11 & 2.01 & \textbf{1.32} & 9748.29 & 5258.08 & 4574.76 & \textbf{4379.26} \\
        \midrule
        \multirow{3}{*}{Training}
        & Meta-240 & 32,768 & (32, 32) &106.19 & 19.54 & 10.71 & \textbf{5.63} & 5228.45 & 11169.38 & \textbf{2561.15} & 3692.80 \\
        & Meta-480 & 32,768 & (32, 32) &62.83 & 12.59 & 8.35 & \textbf{5.25} & 6404.93 & 18027.10 & \textbf{3079.37} & 4210.19 \\
        & Meta-788 & 32,768 & (32, 32) &76.91 & 14.76 & 9.15 & \textbf{4.65} & 10069.52 & 21967.18 & \textbf{5008.38} & 6131.40 \\
        \bottomrule
    \end{tabular}
\end{table*}

\subsection{End-to-End Latency Breakdown Including Index Grouping}
\begin{figure}[ht]
    \centering
    \includegraphics[width=0.8\columnwidth]{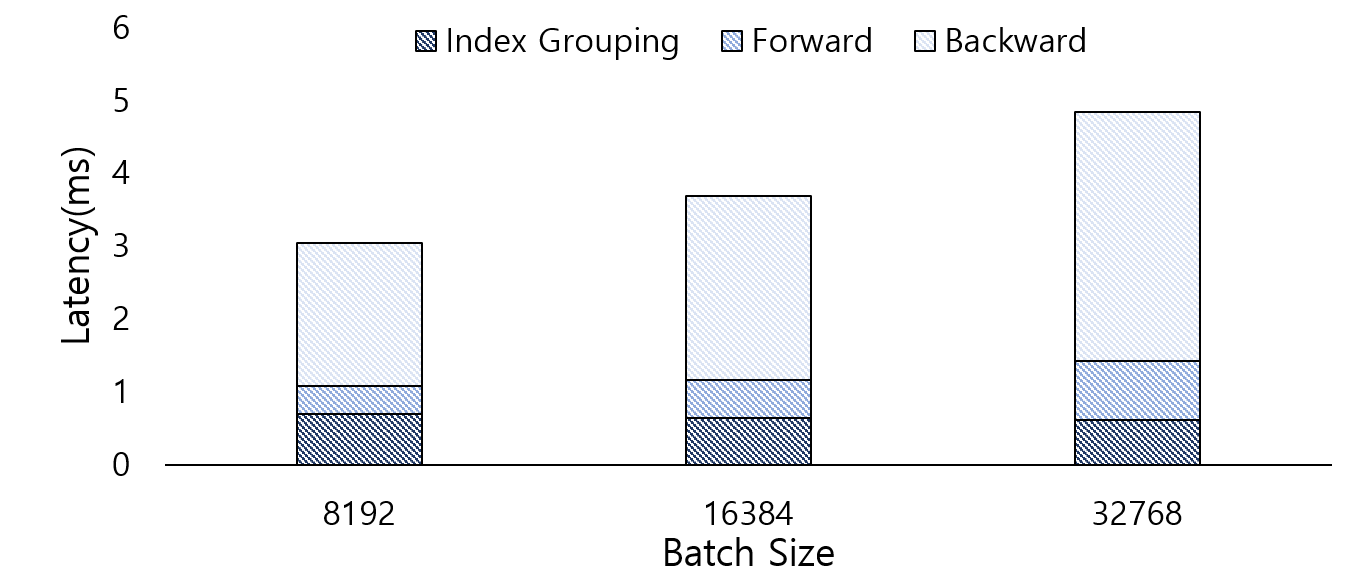}
    \caption{End-to-end training latency breakdown across varying batch sizes on the Meta-788 dataset, including prefix-based index grouping, forward execution, and backward execution.
    %End-to-end training latency breakdown. Breakdown of end-to-end training latency across varying batch sizes on the Meta-788 dataset, detailing the execution times of prefix-based index grouping, forward, and backward execution.
    }
    \label{fig:e2e_breakdown}
\end{figure}

Figure~\ref{fig:e2e_breakdown} presents the end-to-end training latency of FlowTT, decomposed into index grouping, forward, and backward execution. The index grouping stage includes input index sorting, exact unique extraction, mixed-radix conversion, prefix extraction, and RLE.
While index grouping accounts for a noticeable fraction of the total latency at small batch sizes, its relative contribution steadily diminishes with larger batches, and the backward pass remains the dominant contributor across all batch sizes. This indicates that the preprocessing overhead introduced by prefix-based grouping does not scale proportionally with the batch size and is therefore effectively amortized. Meanwhile, as the batch size increases, more inputs share common prefixes, which reduces redundant computations within both forward and backward kernel executions even as their absolute execution times grow. This confirms that the benefits of prefix-based reuse persist at the system level and remain effective across the entire execution pipeline.

\subsection{Ablation of Dynamic Scheduling and Inter-Block Reuse}
To isolate the contributions of FlowTT’s scheduling design, we compare three variants: (i) static prefix mapping, (ii) PT with Work Stealing (WS), and (iii) PT+WS with L2 checkpointing. 
Table~\ref{tab:schedule_ablation} summarizes representative results measured on Meta-788 with a batch size of 32,768. Latency is normalized to that of static prefix mapping, which is set to 1.00.
Static prefix mapping maximizes block-local reuse within each prefix but suffers from severe load imbalance due to variation in prefix group sizes. As a result, some blocks become idle early, and the overall execution time is dominated by the longest prefix group, leading to higher latency.
Introducing PT and WS effectively mitigates this imbalance. Blocks that finish early dynamically steal remaining chunks from unfinished groups, reducing idle time and improving overall utilization. Consequently, the normalized latency decreases from 1.00 to 0.87, indicating that dynamic scheduling is a dominant factor in performance improvement.
Adding L2 checkpointing further reduces latency to 0.85 by recovering reuse that would otherwise be lost during work stealing. When work migrates across blocks, prefix-level intermediate results can be reused via the L2 cache, thereby avoiding redundant recomputation.
These results show that the performance gains of FlowTT stem not only from improved load balancing but also from preserving reuse under irregular execution patterns. 
%Since this mechanism directly addresses the inherent irregularity of TT gather, the scheduling effects can be clearly isolated and interpreted in the inference setting.

\begin{table}[!t]
    \centering
    \caption{Ablation results of dynamic scheduling and inter-block reuse on Meta-788 with batch size 32,768. Latency is normalized to static prefix mapping (set to 1.00), and lower values indicate better performance.}
    \label{tab:schedule_ablation}
    \small
    \setlength{\tabcolsep}{6pt}
    \begin{tabular}{lc}
        \toprule
        Scheme  & Normalized Latency $\downarrow$ \\
        \midrule
        Static prefix mapping    & 1.00 \\
        PT + WS                  & 0.87 \\
        PT + WS + L2 checkpoint  & \textbf{0.85} \\
        \bottomrule
    \end{tabular}
\end{table}

\subsection{Ablation of Prefix-Based Reuse over Exact Deduplication}
\begin{figure}[t]
    \centering
    \includegraphics[width=0.8\columnwidth]{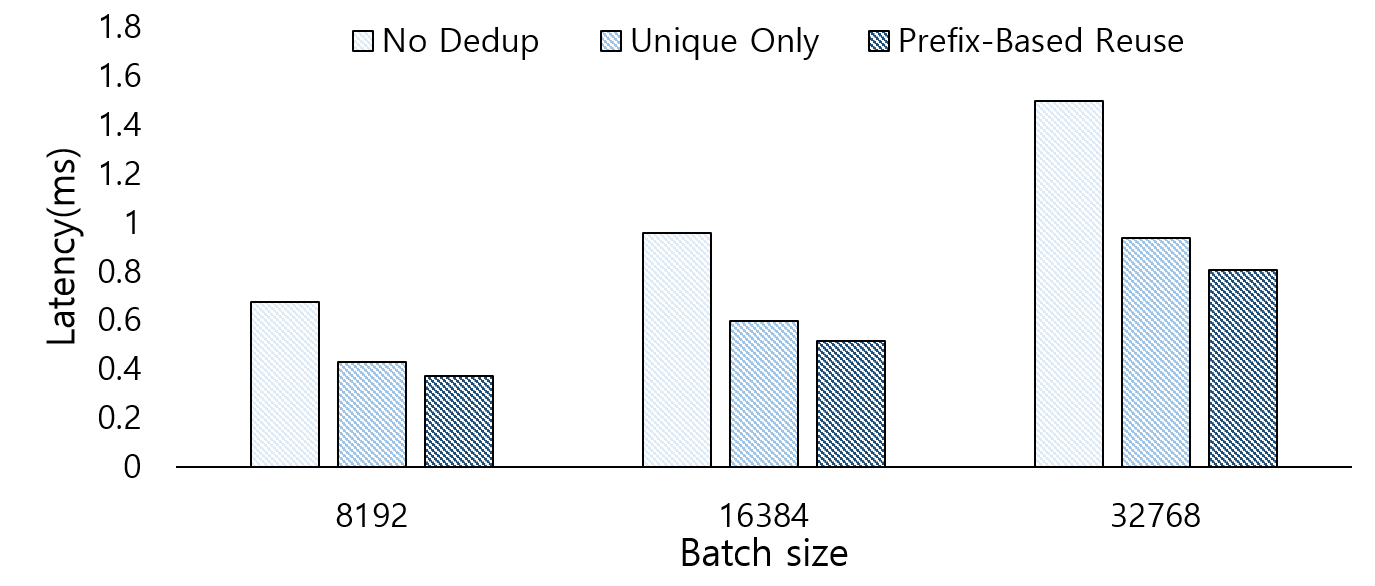}
    \caption{Ablation of prefix-based reuse across varying batch sizes. Compared to exact deduplication (Unique Only), FlowTT’s Prefix-Based Reuse further reduces latency by exploiting shared computation among inputs with common prefixes.}
    \label{fig:reuse_ablation}
\end{figure}

Figure~\ref{fig:reuse_ablation} compares the latency of three configurations to isolate the effect of prefix-based reuse in FlowTT: (i) No Dedup, which performs no deduplication, (ii) Unique Only, which removes only identical embedding indices, and (iii) Prefix-Based Reuse, which combines exact deduplication with prefix-based grouping.
Comparing No Dedup and Unique Only, removing duplicate indices significantly reduces latency by eliminating redundant lookups of identical embeddings. This highlights the impact of repeated memory accesses on the overall computation cost. However, Unique Only does not eliminate redundant computation across different indices that share common higher-order TT core paths, limiting its performance gains.
In contrast, Prefix-Based Reuse extends exact deduplication by grouping inputs that share the same prefix and reusing intermediate results stored in shared memory. By exploiting shared computation flows across different indices, it eliminates additional redundant operations that cannot be addressed by Unique Only. As a result, Prefix-Based Reuse consistently achieves lower latency than Unique Only across all batch sizes, as shown in Figure~\ref{fig:reuse_ablation}.
Furthermore, the performance gap widens as the batch size increases. This is because larger batches expose more inputs sharing common prefixes, amplifying the effectiveness of reuse.
These results demonstrate that exact deduplication alone is insufficient to fully eliminate redundancy in TT gather, and that prefix-level computation flow reuse is essential for achieving further performance improvements.
Notably, prefix sharing is not an artifact of the synthetic benchmark: on full real recommendation traces (Amazon CDs, Amazon Video Games, and MovieLens 25M), the reuse ratio of unique embedding vectors ranges from 0.22 to 0.98, the average prefix group contains 11--28 indices, and 89--96\% of stage-1 TT-core work is reusable through prefix grouping.

\section{Conclusion}
\noindent
% In this paper, we revisited the inefficiency of TT-based embedding lookup not as a sequence of independent GEMM optimizations, but as a failure to exploit computation flows shared across inputs. We proposed FlowTT, a flow-aware execution framework that combines prefix-based index grouping, a fused TT-embedding kernel with on-chip intermediate retention, and dynamic scheduling with work stealing and L2-mediated reuse.
% By aligning task formation, data reuse, and scheduling with the structure of TT gather, FlowTT reduces redundant computation and minimizes off-chip data movement. Experimental results show that FlowTT consistently achieves lower latency and lower peak memory usage than prior TT-based implementations, with increasing benefits at larger batch sizes and table scales.
% These results demonstrate that execution-level optimization, when coupled with compression techniques such as TTD, is critical for realizing practical performance gains in large-scale recommendation systems. Future work includes extending FlowTT to deeper TT decompositions, multi-GPU training environments, and full integration with end-to-end DLRM pipelines.
We presented FlowTT, a flow-aware execution framework for TT-based embedding that redefines TT gather as a prefix-shared irregular computation problem rather than a sequence of independent GEMMs. FlowTT makes reusable computation explicit through prefix-based index grouping, executes TT contractions in a fused on-chip manner, and sustains high utilization under skewed workloads via persistent threads, chunk-based work stealing, and L2 checkpointing. This co-design reduces redundant TT-core computation and global-memory traffic while preserving reuse under irregular execution. On Meta-240, Meta-480, and Meta-788, FlowTT consistently delivered the lowest latency among the compared TT-based baselines. At batch size $32{,}768$, it achieved up to $42.2\%$ lower inference latency and $49.2\%$ lower training latency than EcoRec, and it also attained the lowest inference peak memory usage. Although training requires an additional gradient-aggregation buffer, FlowTT still provides the best latency across all evaluated settings. Overall, the results indicate that the main performance opportunity in TT embedding lies not only in compression itself, but also in how shared computation flows are exposed and mapped onto the GPU memory hierarchy. Future work includes extending FlowTT to deeper TT decompositions and larger-scale multi-GPU DLRM deployments.

\begin{acks}
%This work was partly supported by the Institute of Information \& Communications Technology Planning \& Evaluation (IITP) grants funded by the Korean government (MSIT) (RS-2025-02314443, Development of Flash Memory-based AI Processing Unit for On-Device AI), and the Korea Evaluation Institute of Industrial Technology (KEIT) grant funded by the Korean government (MOTIE) (RS-2026-25534585, 2026 HRD Program for Industrial Innovation).
This work was supported by Institute of Information \& Communications Technology Planning \& Evaluation (IITP) grants funded by the Korea government (MSIT) (RS-2025-02314443, Development of Flash Memory-based AI Processing Unit for On-Device AI; and IITP-2026-RS-2023-00253914, artificial intelligence semiconductor support program to nurture the best talents).
\end{acks}

\bibliographystyle{ACM-Reference-Format}
\bibliography{reference}

@article{naumov2019deep,
  title={Deep learning recommendation model for personalization and recommendation systems},
  author={Naumov, Maxim and Mudigere, Dheevatsa and Shi, Hao-Jun Michael and Huang, Jianyu and Sundaraman, Narayanan and Park, Jongsoo and Wang, Xiaodong and Gupta, Udit and Wu, Carole-Jean and Azzolini, Alisson G and others},
  journal={arXiv preprint arXiv:1906.00091},
  year={2019}
}

@inproceedings{mudigere2022software,
  title={Software-hardware co-design for fast and scalable training of deep learning recommendation models},
  author={Mudigere, Dheevatsa and Hao, Yuchen and Huang, Jianyu and Jia, Zhihao and Tulloch, Andrew and Sridharan, Srinivas and Liu, Xing and Ozdal, Mustafa and Nie, Jade and Park, Jongsoo and others},
  booktitle={Proceedings of the 49th Annual International Symposium on Computer Architecture},
  pages={993--1011},
  year={2022}
}

@article{oseledets2011tensor,
  title={{Tensor-Train} decomposition},
  author={Oseledets, Ivan V},
  journal={SIAM Journal on Scientific Computing},
  volume={33},
  number={5},
  pages={2295--2317},
  year={2011},
  publisher={SIAM}
}

@inproceedings{gupta2012study,
  title={A study of persistent threads style {GPU} programming for {GPGPU} workloads},
  author={Gupta, Kshitij and Stuart, Jeff A and Owens, John D},
  booktitle={2012 Innovative Parallel Computing (InPar)},
  pages={1--14},
  year={2012},
  organization={IEEE}
}

@manual{nvidia_cublas,
  author       = {{NVIDIA Corporation}},
  title        = {{cuBLAS} Library User Guide},
  year         = {2025},
  note         = {CUDA Toolkit 12.8},
  url          = {https://docs.nvidia.com/cuda/cublas/index.html},
  organization = {NVIDIA Corporation}
}

@inproceedings{tzeng2010task,
  title={Task management for irregular-parallel workloads on the {GPU}},
  author={Tzeng, Stanley and Patney, Anjul and Owens, John D},
  booktitle={Proceedings of the Conference on High Performance Graphics},
  pages={29--37},
  year={2010}
}

@inproceedings{shi2020compositional,
  title={Compositional embeddings using complementary partitions for memory-efficient recommendation systems},
  author={Shi, Hao-Jun Michael and Mudigere, Dheevatsa and Naumov, Maxim and Yang, Jiyan},
  booktitle={Proceedings of the 26th ACM SIGKDD International Conference on Knowledge Discovery \& Data Mining},
  pages={165--175},
  year={2020}
}

@inproceedings{acun2021understanding,
  title={Understanding training efficiency of deep learning recommendation models at scale},
  author={Acun, Bilge and Murphy, Matthew and Wang, Xiaodong and Nie, Jade and Wu, Carole-Jean and Hazelwood, Kim},
  booktitle={2021 IEEE International Symposium on High-Performance Computer Architecture (HPCA)},
  pages={802--814},
  year={2021},
  organization={IEEE}
}

@inproceedings{yin2021tt,
  title={{TT-Rec}: {Tensor} train compression for deep learning recommendation models},
  author={Yin, Chunxing and Acun, Bilge and Wu, Carole-Jean and Liu, Xing},
  booktitle={Proceedings of Machine Learning and Systems},
  volume={3},
  pages={448--462},
  year={2021}
}

@inproceedings{wang2022rec,
  title={{EL-Rec}: Efficient large-scale recommendation model training via tensor-train embedding table},
  author={Wang, Zheng and Wang, Yuke and Feng, Boyuan and Mudigere, Dheevatsa and Muthiah, Bharath and Ding, Yufei},
  booktitle={SC22: International Conference for High Performance Computing, Networking, Storage and Analysis},
  pages={1--14},
  year={2022},
  organization={IEEE}
}

@article{adnan2021accelerating,
  title={Accelerating recommendation system training by leveraging popular choices},
  author={Adnan, Muhammad and Maboud, Yassaman Ebrahimzadeh and Mahajan, Divya and Nair, Prashant J},
  journal={Proceedings of the VLDB Endowment},
  volume={15},
  number={1},
  pages={127--140},
  year={2021}
}

@inproceedings{wang2024accelerating,
  title={Accelerating distributed {DLRM} training with optimized {TT} decomposition and micro-batching},
  author={Wang, Weihu and Xia, Yaqi and Yang, Donglin and Zhou, Xiaobo and Cheng, Dazhao},
  booktitle={SC24: International Conference for High Performance Computing, Networking, Storage and Analysis},
  pages={1--15},
  year={2024},
  organization={IEEE}
}

@article{zhang2025flashgemm,
  title={{FlashGEMM}: Optimizing Sequences of Matrix Multiplication by Exploiting Data Reuse on {CPUs}},
  author={Zhang, Junwen and Yang, Weiling and Fang, Jianbin and Dong, Dezun and Chen, Xianzhang},
  journal={ACM Transactions on Architecture and Code Optimization},
  year={2025},
  doi={10.1145/3760784},
  publisher={ACM New York, NY}
}

@inproceedings{dao2022flashattention,
  title={{FlashAttention}: Fast and memory-efficient exact attention with {IO}-awareness},
  author={Dao, Tri and Fu, Dan and Ermon, Stefano and Rudra, Atri and R{\'e}, Christopher},
  booktitle={Advances in Neural Information Processing Systems},
  volume={35},
  pages={16344--16359},
  year={2022}
}

@inproceedings{dao2024flashattention2,
  title={{FlashAttention-2}: Faster attention with better parallelism and work partitioning},
  author={Dao, Tri},
  booktitle={The Twelfth International Conference on Learning Representations (ICLR)},
  year={2024}
}

@inproceedings{shah2024flashattention,
  title={{FlashAttention-3}: Fast and accurate attention with asynchrony and low-precision},
  author={Shah, Jay and Bikshandi, Ganesh and Zhang, Ying and Thakkar, Vijay and Ramani, Pradeep and Dao, Tri},
  booktitle={Advances in Neural Information Processing Systems},
  volume={37},
  pages={68658--68685},
  year={2024}
}

@inproceedings{sethi2022recshard,
  title={{RecShard}: statistical feature-based memory optimization for industry-scale neural recommendation},
  author={Sethi, Geet and Acun, Bilge and Agarwal, Niket and Kozyrakis, Christos and Trippel, Caroline and Wu, Carole-Jean},
  booktitle={Proceedings of the 27th ACM International Conference on Architectural Support for Programming Languages and Operating Systems},
  pages={344--358},
  year={2022}
}

@misc{facebook_fbtt_embedding,
  author       = {{Facebook AI Research}},
  title        = {{FBTT-Embedding}: {Tensor} {Train} based compression library for sparse embedding tables},
  howpublished = {\url{https://github.com/facebookresearch/FBTT-Embedding}},
  year         = {2021},
  note         = {GitHub repository. Accessed: 2025-11-19}
}

@inproceedings{desai2022trade,
  title={The trade-offs of model size in large recommendation models: {100GB} to {10MB} {Criteo-tb} {DLRM} model},
  author={Desai, Aditya and Shrivastava, Anshumali},
  booktitle={Advances in Neural Information Processing Systems},
  volume={35},
  pages={33961--33972},
  year={2022}
}

@inproceedings{aila2009understanding,
  title={Understanding the efficiency of ray traversal on {GPUs}},
  author={Aila, Timo and Laine, Samuli},
  booktitle={Proceedings of the conference on high performance graphics 2009},
  pages={145--149},
  year={2009}
}

@inproceedings{chatterjee2011dynamic,
  title={Dynamic task parallelism with a {GPU} work-stealing runtime system},
  author={Chatterjee, Sanjay and Grossman, Max and Sb{\^\i}rlea, Alina and Sarkar, Vivek},
  booktitle={International Workshop on Languages and Compilers for Parallel Computing},
  pages={203--217},
  year={2011},
  organization={Springer}
}

@inproceedings{zha2022autoshard,
  title={{AutoShard}: Automated embedding table sharding for recommender systems},
  author={Zha, Daochen and Feng, Louis and Bhushanam, Bhargav and Choudhary, Dhruv and Nie, Jade and Tian, Yuandong and Chae, Jay and Ma, Yinbin and Kejariwal, Arun and Hu, Xia},
  booktitle={Proceedings of the 28th ACM SIGKDD Conference on Knowledge Discovery and Data Mining},
  pages={4461--4471},
  year={2022}
}

@inproceedings{agarwal2023bagpipe,
  title={{Bagpipe}: Accelerating deep recommendation model training},
  author={Agarwal, Saurabh and Yan, Chengpo and Zhang, Ziyi and Venkataraman, Shivaram},
  booktitle={Proceedings of the 29th Symposium on Operating Systems Principles},
  pages={348--363},
  year={2023}
}

@misc{meta_dlrm_datasets_2022,
  author = {{Meta Research}},
  title = {{DLRM} Datasets: Synthetic Embedding Bag Data},
  howpublished = {\url{https://github.com/facebookresearch/dlrm_datasets/tree/main/embedding_bag/2022}},
  year = {2022},
  note = {GitHub repository},
  organization = {Meta Platforms}
}

@inproceedings{ginart2021mixed,
  title={Mixed dimension embeddings with application to memory-efficient recommendation systems},
  author={Ginart, Antonio A and Naumov, Maxim and Mudigere, Dheevatsa and Yang, Jiyan and Zou, James},
  booktitle={2021 IEEE International symposium on information theory (ISIT)},
  pages={2786--2791},
  year={2021},
  organization={IEEE}
}


% \appendix









\end{document}